\documentclass[conference]{IEEEtran}
\IEEEoverridecommandlockouts
\usepackage{cite}
\usepackage{amsmath,amssymb,amsfonts}
\usepackage{algorithmic}
\usepackage{graphicx}
\usepackage{textcomp}
\usepackage{xcolor}
\usepackage{enumitem}

\def\BibTeX{{\rm B\kern-.05em{\sc i\kern-.025em b}\kern-.08em
    T\kern-.1667em\lower.7ex\hbox{E}\kern-.125emX}}
\begin{document}

\title{A Dynamic Resource Scheduling Algorithm Based on Traffic Prediction for Coexistence of eMBB and Random Arrival URLLC \\

}

\author{
    \IEEEauthorblockN{Yizhou Jiang\IEEEauthorrefmark{1}\IEEEauthorrefmark{2} Xiujun Zhang\IEEEauthorrefmark{2} Xiaofeng Zhong\IEEEauthorrefmark{1}\IEEEauthorrefmark{3} Shidong Zhou\IEEEauthorrefmark{1}\IEEEauthorrefmark{3}}
    \IEEEauthorblockA{\IEEEauthorrefmark{1}Department of Electronic Engineering, Tsinghua University, Beijing, China}
    \IEEEauthorblockA{\IEEEauthorrefmark{2}Beijing National Research Center for Information Science and Technology}
    \IEEEauthorblockA{\IEEEauthorrefmark{3}State Key Laboratory of Space Network and Communications\\
    Emails: jiangyz19@mails.tsinghua.edu.cn, \{zhangxiujun, zhongxf, zhousd\}@tsinghua.edu.cn}
}
\maketitle

\begin{abstract}

    In this paper, we propose a joint design for the coexistence of enhanced mobile broadband (eMBB) and ultra-reliable and random low-latency communication (URLLC) with different transmission time intervals (TTI): an eMBB scheduler operating at the beginning of each eMBB TTI to decide the coding redundancy of eMBB code blocks, and a URLLC scheduler at the beginning of each mini-slot to perform immediate preemption to ensure that the randomly arriving URLLC traffic is allocated with enough radio resource and the eMBB traffic keeps acceptable one-shot transmission successful probability and throughput.
    The framework for schedulers under hybrid-TTI is developed and a method to configure eMBB code block based on URLLC traffic arrival prediction is implemented. Simulations show that our work improves the throughput of eMBB traffic without sacrificing the reliablity while supporting randomly arriving URLLC traffic.

\end{abstract}


\section{Introduction}

As mobile Internet evolves, the scenarios of mobile application become more diversified. In the fifth generation communication system New Radio (5G-NR). Three major scenarios are defined, which are enhanced Mobile Broadband (eMBB), Ultra-Reliable Low Latency Communications (URLLC), and massive Machine Type Communications (mMTC) corresponding to different key performance indicators (KPI)\cite{vision6g}. eMBB services require very large throughput, while URLLC services require strict latency and high reliability, and mMTC services have to handle massive access of Internet-of-Things (IoT) devices\cite{popovski20185g}.

To meet the diversified traffic requirements, the wireless communication systems have to introduce a more sophisticated design of schedule mechanisms. For example, when multiplexing eMBB and URLLC in orthodox frequency division multiplexing (OFDM) systems, these two kinds of services are suitable for different transmission time intervals (TTI), and in general, a scheme called hybrid TTI is adopted\cite{hybridTTI}. To achieve higher spectral efficiency, gain long-code benefits, and reduce signaling overhead, a longer TTI is adopted by eMBB transmission, which can be called eMBB TTI; while a shorter TTI is adopted by URLLC transmission, which can be called eMBB TTI, to attain millisecond-level response time. In the third generation partnership project (3GPP) Release-16, introduce mini-slot transmission mechanism to better support low-latency transmission\cite{g38824}. A traditional slot consists of 12 or 14 OFDM symbols, while a mini-slot can contain as few as one OFDM symbol\cite{g38211}.

Many efforts have been done about the realization of hybrid TTI machanism by both academia and industry. 
To dynamically multiplex eMBB and URLLC traffic, two approaches, preemption and superposition, have been suggested in \cite{intel}. Preemption scheme is preferred for almost no interference to URLLC traffic to achieve high reliability in downlink scenario.
Preemption/puncture scheme is applied in 
\cite{pedersen_punctured_2017}\cite{li_deep_2020}\cite{yin_multiplexing_2020}\cite{alsenwi_intelligent_2021}\cite{hsu_embb_2022}\cite{zhao_resource_2022} to implement eMBB and URLLC multiplexing. When an urgent URLLC transmission occurs but all resource blocks (RB) are allocated to eMBB transmissions, signals of some ongoing eMBB transmissions have to be nulled in a mini-slot,and then these resource will be re-allocated to the URLLC transmission and be overwritten by URLLC symbols to meet the stringent latency constraint. The common limitations of these work are that a linear loss model\cite{anandJointSchedulingURLLC2020} are utilized to assess the throughput loss of code blocks, i.e. the amount of successful throughput are proportional to the resource not preempted. This linear loss model is a conclusion of the channel capacity of AWGN channel with erasures\cite{ErasureChannel}. 
Although this model can optimize the long-term average throughput which can be implemented by hybrid automatic repeat request (HARQ)\cite{ahmed2021hybrid}, but can not ensure the BLER in one-shot transmission. Because the preemption will increase the actual code rate of eMBB code block, leading to the degradation of BLER and the re-transmission will increase the latency of eMBB transmission, both of which are also pursued in the actual communication systems.
As a result, these work failed to meet the performance requirements of real-world eMBB services.

To address this issue, \cite{huang_deluxe_2022} jointly optimizes the throughput and the reliability of eMBB traffic, by assuming that the exact amount of future arrival URLLC packets in the next eMBB TTI is known before the eMBB TTI starts.
However, when the URLLC traffic arrives randomly\cite{popovski2018wireless}, this amount could not be available in advance, therefore new scheculing method need to be developed. 

Based on the weakness of the above works, this paper propose a schedule framework that can ensure both throughput and reliability of eMBB transmission under random arrivals of URLLC traffic. In our work, the arrivals of URLLC traffic are not necessary before a TTI of eMBB transmission, while only the statistics of URLLC are required to optimize the allocation and configuration of eMBB transmission to achieve both throughput and reliability requirements.

\section{System Model}
A single cell downlink OFDM system is considered, where a single base station (BS) serves $K$ UEs, 
including $N_e$ eMBB users and $N_u$ URLLC users. This model can also be easily generalized to the multi-cell downlink scenario considering inter-cell interference. The total system bandwidth is $W$.

\subsection{Time-Frequency Resource Granularity}
\begin{figure}[h]
    \centering
    \includegraphics[width=\linewidth]{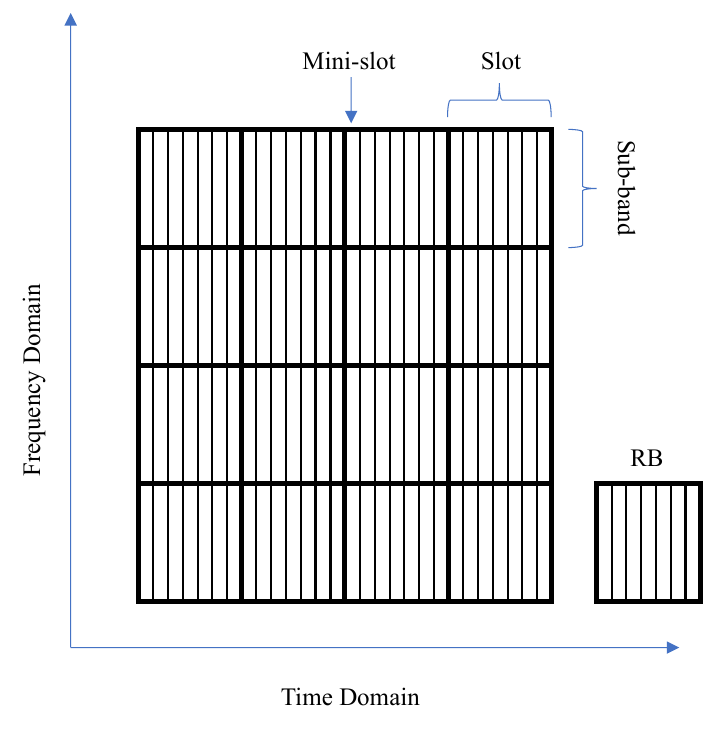}
    \caption{Time-Frequency Resource Granularity}
    \label{fig:tfr}
\end{figure}

Radio resource is divided in to sub-bands in frequency domain, and (mini-)slot in time domain as shown in Fig.\ref{fig:tfr}

In frequency domain, the total bandwidth $W$ is equally divided into $N_s$ sub-bands, size of which is $B=W/{N_s}$. A sub-band is the minimum schedule granularity of both kinds of traffic. Each sub-band consists of $N_{SS}$ sub-carriers.

Based on hybrid-TTI scheme, two schedule granularities are applied to eMBB traffic and URLLC traffic respectively: an eMBB TTI, consisting of $T_s=M\times L_m$ OFDM symbols; a URLLC TTI consisting of $L_m$ OFDM symbols, which is obtained by dividing an eMBB TTI into $M$ equal parts. For simplicity, we will refer to them respectively as \emph{slot} and \emph{mini-slot} in the following text.

\subsection{Channel Model and CSI Knowledge}
A frequency selective channel is adopted. Let $\mathbf{h}_{i,k}$ denotes the channel of the eMBB user $i$ on sub-band $k$. And the channel is constant within a slot.
We also assume that the BS acquires the eMBB UEs' CSI knowledge, but no URLLC UEs' CSI knowledge, because the instantaneous CSI is usually available for URLLC traffic\cite{perez2022robust}.

\subsection{Traffic Model}
Different traffic arrival models are applied to two kinds of services.

\textbf{eMBB traffic:} full-buffer model. There are always data can be scheduled and transmitted to every eMBB user.

\textbf{URLLC traffic:} Poisson model. The number of URLLC packets generated per mini-slot is independent and follows Poisson distribution with an average arrival rate of $\lambda$ packets per mini-slot\cite{abreu2019system}. Every URLLC packet occupies same amount of time-frequency resource of a mini-slot in time domain and a sub-band in frequency domain.
We denote the index of a mini-slot as $\tau$, the number of packet arriving at the queue in mini-slot $\tau$ by $G_{\tau}$, the number of packet leaving the queue and being scheduled in mini-slot $\tau$ by $D_{\tau}$, and the length of URLLC queue at the start of mini-slot $\tau$ by $L_{\tau}$.
So, we have the state transition equation,
\begin{equation}
    L_{\tau+1}=L_{\tau}+ G_{\tau}- D_{\tau}.
    \label{eq:qlen}
\end{equation}
And the probability of the number of arrival URLLC packets $G_{\tau}=n$ is
\begin{equation}
    \mathbb{P}(G_{\tau}=n)=\frac{\lambda^n \exp(-\lambda) }{n!}
    \label{eq:g_pmf}
\end{equation}

\subsection{Criteria for Evaluating the Performance}
Different criteria are used to evaluate the performance of these two kinds of services.

\textbf{eMBB services:}
The main criterion to optimize is the total downlink throughput of eMBB users, while the reliability under preemption of these links also have to be ensured. We denote the block error rates (BLER) of eMBB user $i$ as $\epsilon_i$, and it has to be ensured that $\epsilon_i\leq \epsilon$, where $\epsilon$ is the greatest BLER that can be tolerated by eMBB sevices.

\textbf{URLLC services:}
The latency and reliability are two main aspects of URLLC services. Since the focus of this paper is to optimize eMBB performance under preemption, for simplicity, a URLLC transmission is considered successful once scheduled and allocated resource. And the latency is specially referred to the waiting time in the queue to be scheduled.

\subsection{Schedule Rule}
The resource schedule proplem is divided to two time scale, namely slot-level and mini-slot-level for eMBB packets and URLLC packets respectly. A resource block (RB) occupies a slot in time and a sub-band in frequency, and is the fundamental scheduling unit for eMBB services.
\begin{figure*}[!t]
    \centering
    \includegraphics[width=\textwidth]{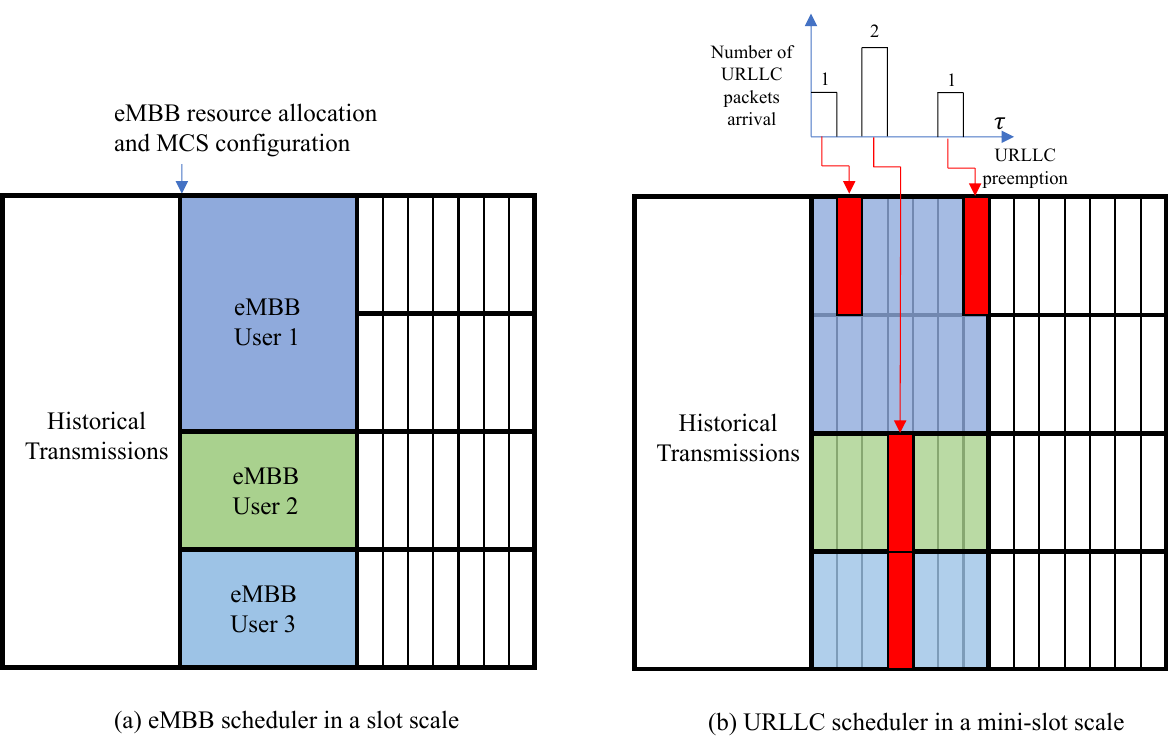}
    \caption{Schedulers in two timescales.}
    \label{fig:wf}
\end{figure*}
\textbf{slot-level:}
At the start of a slot, every RBs will either be allocated to an eMBB user or reserved for URLLC services that will arrive during the slot.
Fig.\ref{fig:wf}(a) shows the allocation of radio resource after slot-level scheculing.
\textbf{mini-slot-level:}
At the start of a mini-slot, several packets whose number is denoted as $D$ in the URLLC queue will be scheduled and $D$ RBs will be preempted during this mini-slot. Existing symbols of eMBB services in this mini-slot and chosen RBs will be overwritten by URLLC packets.
Fig.\ref{fig:wf}(b) shows the allocation of radio resource after mini-slot-level scheculing.
\subsection{Joint Coding}
An eMBB block is jointly coded and interleaved over the RBs allocated. A proper modulation and code scheme (MCS) will be configured. And part of the coded bits will be randomly dropped because of the preemption by URLLC packets, which will increase the risk of block error. 

\subsection{Receiver}
\textbf{URLLC receiver:} listen to the channel in a period of mini-slot, when a packet transmitting to it is detected, offload and decode the packet.

\textbf{eMBB receiver:} at the end of a slot, BS will inform the eMBB user of the positions of symbols preempted in the OFDM grid, and then eMBB UE will avoid to be mislead by these symbols when decoding. If the remaining channel capacity after being occupied is less than the number of information bits encoded in the code block, decoding fails.

\section{Scheduler Design}
The entire scheduling problem is divided into two phases: (i) eMBB scheduling and (ii) URLLC scheduling.

\subsection{URLLC Scheduler}
The URLLC scheduler needs to decide, at the start of each mini-slot, which sub-band(s) need to be premepted for URLLC packet transmission. In this subsection, we will provide the scheduling framework first and then design the proper scheduling policy, based on the information of the current RB allocation and MCS configuration of eMBB traffic.
\subsubsection{Scheduler Framework}
A URLLC scheduler includes three parts: state, decision and policy function. Input the state of the system into the policy function, then we will get the decision of URLLC scheduling.

The input of the URLLC scheduler is the state of system at current mini-slot $\tau$ denoted by $\mathbf{s}(\tau)$, which includes the slot-level state of eMBB code blocks denoted by $\mathbf{x}(t)$ describing capacity and MCS configuration. And $\mathbf{s}(\tau)$ also includes the mini-slot-level state of the URLLC queue and the ratio of resource preempted of eMBB code blocks, denoted by $\mathbf{z}(\tau)$. 
We have 
\begin{equation}
    \mathbf{s}(\tau)=[\mathbf{x}(t),\mathbf{z}(\tau)]^T.
    \label{eq:state}
\end{equation}
In this paper, we implement a simplified version of state $\mathbf{s}(\tau)$, where $\mathbf{x}(t)=[x_0(t),x_1(t),\dots,x_{N_s-1}(t)]$, $x_k(t)$ denotes the proportional fairness metric of RB indexed by $k$ at slot $t$, and $\mathbf{z}(\tau)=L_{\tau}$ is the length of URLLC queue at mini-slot $\tau$ in \eqref{eq:qlen}.


The output of the URLLC scheduler is the decision of which sub-band(s) need to be premepted for URLLC packet transmission at mini-slot $\tau$. We denote the decision by $\mathbf{\alpha}(\tau)$, whose component $\alpha_k(\tau)$ represents whether the RB indexed by $k$ is preempted in mini-slot $\tau$:
\begin{equation}
\alpha_k(\tau) = 
\begin{cases}
    1, & \text{if } \text{ preempt RB }$k$, \\
    0, & \text{if } \text{ not preempt RB }$k$.
\end{cases}
\label{eq:decision}
\end{equation}

Finally we give the expression of policy function $\pi$:
\begin{align}
    \pi: \mathbb{R}^{N_{s}+1} &\rightarrow \{0,1\}^{N_s}, \\
    \mathbf{s}(\tau) &\mapsto [\alpha_0(\tau), \alpha_1(\tau), \ldots, \alpha_{N_s-1}(\tau)]^T,
\end{align}


\subsubsection{Deterministic Greedy URLLC Scheduler}
Greedy preemption policy combined with proportional fairness criterion proposed in \cite{pedersen_punctured_2017}, where URLLC traffic prioritize preempting eMBB users with poor channel quality, is proved to be efficient for URLLC scheduling. This greedy strategy is applied to our scheduling framework.
\begin{enumerate}[label=(\roman*),left=0pt, labelsep=1em, itemsep=1em]
    \item
Considering the fairness, we adopt proportional fairness criterion to depict the slot-level state of RBs, $\mathbf{x}(t)$. 
If RB $k$ has a higher value of $x_k(t)$, it means that it is more important of this RB and causes greater loss when preempted. From a proportional fairness perspective, this loss should be inversely proportional to the historical average throughput of the user, whom RB $k$ is allocated to, and directly proportional to the user's channel capacity on RB $k$.

To obtain the criterion, we first calculate the historical average throughput $S_i(t)$ for eMBB user $i$ until slot $t$:

\begin{equation}
S_i(t)=
\begin{cases}
    (1-{\eta})S_i(t-1)+{\eta}r_i(t), &\text{if } t>0,\\
    0 , &\text{if } t=0.
\end{cases}
\label{eq:history}
\end{equation}
where $\eta\in (0,1)$ is the aging factor, and $r_i(t)$ is the successful throughput of eMBB user $i$ at slot $t$.

And the metric for RBs to be preempted is
\begin{equation}
x_k(t)=\frac{\log(1+\gamma_{i(k),k}(t))}{S_{i(k)}(t-1)},
\label{eq:metric}
\end{equation}
where RB $k$ is already allocated to eMBB user $i(k)$ at the start of the slot, and $\gamma_{i(k),k}(t)$ is SNR of eMBB user $i(k)$ on RB $k$.
The metric \eqref{eq:metric} is updated at the start of every slot.

\item
Decide $D_{\tau}$, the number of URLLC packets to leave the URLLC queue and to be scheduled in $\tau$ as in \eqref{eq:qlen}. To minimize queuing latency, we adopt a best-effort scheduling strategy, where 
\begin{equation}
D_{\tau}=\min(L_{\tau},N_s).
\label{eq:best_effort}
\end{equation}
If the queue length $L_{\tau}$ is not greater than the number of sub-bands $N_s$, the queue is cleared, otherwise $N_s$ packets leave the queue. And all packets leaving the queue will each preempt a sub-band. 
\item
Denote $N_p(\tau)$ as the amount of sub-bands preempted in mini-slot $\tau$. Based on the best-effort strategy, we have
$$
N_p(\tau)=D_{\tau}.
$$
Preempt $N_p(\tau)$ RBs with the lowest metrics $x_k(t)$, where $Mt\leq \tau \leq (M+1)t-1$ by URLLC preempter $\mathbf{\alpha}(\tau)=\pi(\mathbf{s}(\tau))$, and the components of $\mathbf{\alpha}(\tau)$ is
\begin{equation}
\alpha_i(\tau) =
\begin{cases}
    1, & i \in \{\text{RBs with the} \\
       &        \text{lowest-}N_p(\tau)\text{ metrics}\}, \\
    0, & \text{otherwise.}
\end{cases}
\label{eq:greedy}
\end{equation}

\end{enumerate}

\subsection{eMBB Scheduler and MCS Configuration}
 Since our primary objective is to optimize the code block configuration, we assume that the RB allocation has already been provided by the higher-layer system.

To decide each eMBB code block's coding redundancy, the scheduler have to know not only the channel quality of each eMBB code block, but also \emph{the amount of resources of each eMBB code block that will be preempted during transmission}. For simplicity, we call the amount \emph{preemption pattern}. Actually, based on the distribution of $G_{\tau}$ and the preempting policy $\mathbf{\pi}(\mathbf{x})$, we can calculate the distribution of $D_{\tau}$ and predict the corresponding preemption pattern in a queuing theory method.

By the above results, we can calculate the BLER under specific coding redundancy, and choose a proper coding redundancy for target BLER.

\subsubsection{The Distribution of $N_p(\tau)$}
First we give the state transition equation of the queue length $L_{\tau}$ under best-effort strategy, by substituting \eqref{eq:best_effort} into \eqref{eq:qlen}:
\begin{equation}
    L_{\tau+1}=\max\{L_{\tau}-N_{s},0\}+ G_{\tau},
    \label{eq:qlen2}
\end{equation}

The distribution of $G_{\tau}$ is already given in \eqref{eq:g_pmf}. For simplicity, we denote $a_n=\mathbb{P}(G_{\tau}=n)$. Then we denote $p_{n,\tau}=\mathbb{P}(L_{\tau}=n)$.
We have the equation of probability when $\tau$ is not the last mini-slot of the slot,
\begin{equation}
p_{i,\tau}=a_{i}(p_{0,\tau-1}+...+p_{N_s,\tau-1})+\sum_{k=0}^{i-1}a_{k}p_{N_s+i-k,\tau-1},
\label{eq:L_p}
\end{equation}
and the queue length $L_{\tau}$ is deterministic at the end of a slot, so
\begin{equation}
p_{i,\tau}=\mathbb{I}(i=L_{\tau}).
\label{eq:L_d}
\end{equation}

Then, the distribution of $D_{\tau}$ can be derived from the distribution of $L_{\tau}$. Let $q_{n,\tau}=\mathbb{P}(G_{\tau}=n)$, and we have
\begin{equation}
    q_{n,\tau}=
    \begin{cases}
        p_{n,\tau}, & \text{for}\quad 0\leq n<N_s, \\
        \sum_{n=N_s}^{\infty}p_{n,\tau}, & \text{for}\quad n=N_s. 
    \end{cases}
    \label{eq:G_p}
\end{equation}
Since $N_p(\tau)=D_{\tau}$, we get the distribution of the total amount of preemption.
\subsubsection{Preemption Pattern}
Preemption pattern is necessary for BLER prediction. Since the URLLC scheduler always preempt the RBs with the lowest-$D_{\tau}$ metric $x_k(t)$, we propose a sorting-based computation method for preemption pattern.

First, at the start of slot $t$, sort the RBs indexed by $k=0,\dots,N_s-1$ in ascending order according to metric $x_k(t)$. We get a temporary index $[k]=0,\dots,N_s-1$ for RBs after sorting, where
\begin{equation}
x_{[0]}(t)\leq x_{[1]}(t) \leq \dots \leq x_{[N_s]}(t).
\label{eq:sort}
\end{equation}
If $D_{\tau}=0$, no RBs are preempted; if $D_{\tau}=1$, RB $[0]$ is preempted; if $D_{\tau}=n$, from RB $[0]$ to RB $[n-1]$ are preempted.

So, the number of times RB $[k]$ is preempted at mini-slot $\tau$ is
temporary index $[k]=0,\dots,N_s-1$ for RBs after sorting, where
\begin{equation}
\mathbb{I}(D_{\tau}> k)=
\begin{cases}
0, & \text{if } D_{\tau}\leq k, \\
1, & \text{if } D_{\tau}> k.
\end{cases}
\label{eq:N_preempted_minislot}
\end{equation}
And the total number of times RB $[k]$ is preempted at slot $t$ is
\begin{equation}
y_{[k]}=\sum_{\tau=Mt}^{M(t+1)-1}\mathbb{I}(D_{\tau}> k).
\label{eq:N_preempted_slot}
\end{equation}
\subsubsection{BLER Prediction}
Threshold model is mentioned in \cite{anandJointSchedulingURLLC2020}. Referring to this work,
we propose a threshold-like model to predict BLER and the threshold is the coding redundancy. If the preempted ratio of a RB is less than the coding redundancy, the BLER is approximately $0$, otherwise is approximately $1$.The BLER for RB $k$ is
\begin{equation}
f_k(y_k)=\frac{1}{1+\exp(-\mu_k(y_k/M-\theta_k))},
\label{eq:sigmoid}
\end{equation}
which is a sigmoid function, where $\theta_k$ is the threshold and $\mu_k$ is a hyperparameter to adjust the steepness around the threshold.

\section{Performance}
\subsection{Compared Methods}
We compare our method with separate bands\cite{abreu2019multiplexing} as a baseline and the scenario where the future URLLC event is known before eMBB scheduling as the upper bound.
\subsubsection{Separate Bands Baseline}
We compare this method as a baseline. In this method, the sub-bands are divided into two kinds staticly and exclusively in a slot: one for eMBB transmission, and the other for URLLC transmission. So in this method, different services are transmitted over different sub-bands independently.
Before the eMBB scheduling, we take expectation on the amount of URLLC packets, and reserve enough sub-bands for them, such that the total average bandwidth demand of URLLC is not greater the bandwidth allocated exclusively to them. The rest sub-bands can be allocated to eMBB users, and conduct user selection based on a proportional-fairness criterion.
\subsubsection{Upper Bound}
In this scenario, the event of URLLC arrival in the next slot is already known before eMBB scheduling, and we can pre-puncture in the OFDM grid. Then, the rest radio resource can be allocated to eMBB users. Since the arrival event in the future is foreknown, the uncertainty in the arrival of URLLC will be excluded. This scenario is compared as the performance upper bound.

\subsection{Simulation Settings}
In the next simulation, total system bandwidth is $B=9MHz$, and the number of sub-bands or RBs is $N_s=25$, so the bandwidth of each sub-band is $B=W/{N_s}=360kHz$. Each sub-band consists of $N_{SS}=12$ sub-carriers, and the sub-carrier spacing (SCS) is $30kHz$.

In 5G numerology, the SCS of $30kHz$ corresponds to a length of slot $0.5ms$, which consists of $T_s=14$ OFDM symbols. The length of mini-slot is $L_m=2$ OFDM symbols, so the number of mini-slot in a slot is $M=T_s/L_m=7$.

The number of eMBB users is $N_{eMBB}=25$. The arrival rate of URLLC in BS side $\lambda$ is varies from $0.0$ to $2.5$ per mini-slot. The target BLER of eMBB services is $\epsilon=10\%$. The number of slot we have performed $10,000$ simulations for each value of $\mu$. 

\subsection{Results}
\begin{figure}[h]
    \centering
    \includegraphics[width=\linewidth]{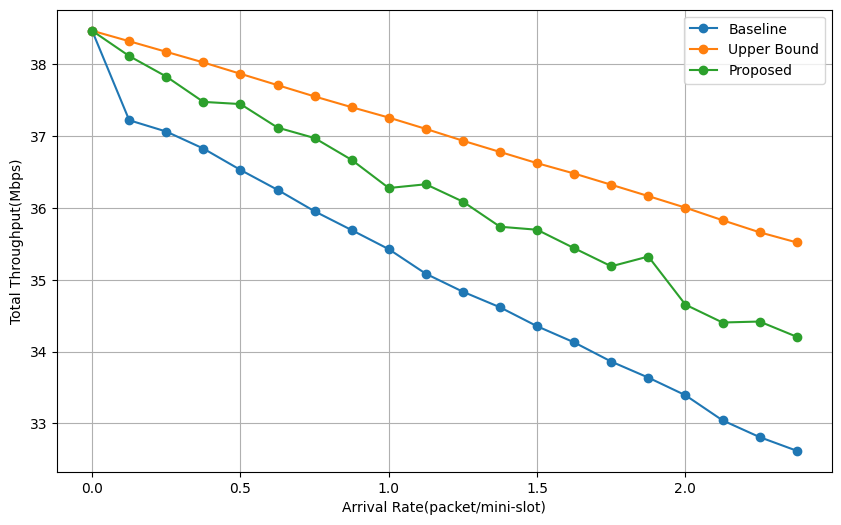}
    \caption{Throughput vs URLLC arrival rate for three methods}
    \label{fig:thp}
\end{figure}

\begin{figure}[h]
    \centering
    \includegraphics[width=\linewidth]{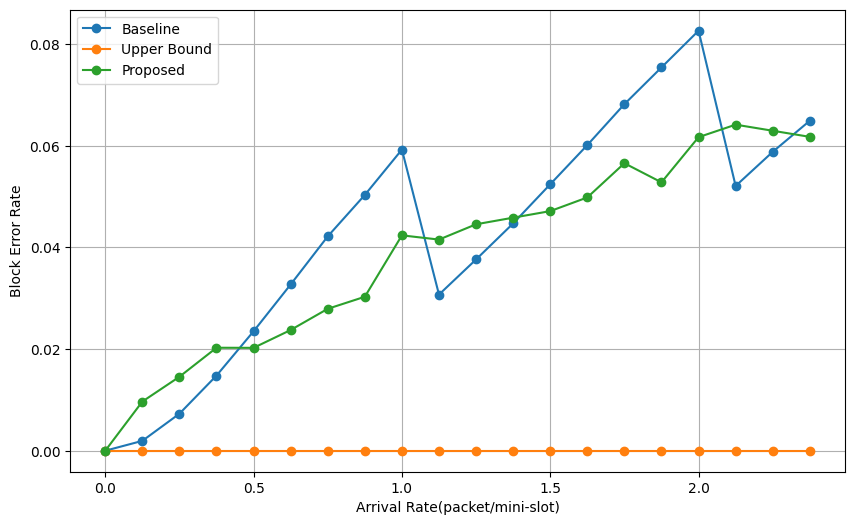}
    \caption{BLER vs URLLC arrival rate for three methods}
    \label{fig:ble}
\end{figure}
Fig.\ref{fig:thp} shows the throughput of the three methods. The proposed method (green line) can achieve approximately $98\%$ throughput of the upper bound (orange line). The fluctuation of URLLC arrival causes the gap between the proposed method and the upper bound. Comparing with the baseline (blue line), the proposed method gains $5\%$ increase, since in some mini-slot when the realization of URLLC arrival is less than the expectation, some sub-bands is not allocated to any eMBB or URLLC users, so the radio resource is left idle.

Fig.\ref{fig:ble} shows the BLER of the three methods. The proposed method (green line) can achieve lower BLER than the baseline (blue line) at most value of URLLC arrival rate $\lambda$, especially when the average URLLC arrival rate $\lambda$ is slightly below but close to a interger. This is because that the amount of sub-bands is discrete, if $\lambda$ is slightly below but close to a interger, then it has a higher probability for the number of arrival URLLC packets exceeds the number of sub-bands reserved and causes more block error. Since in the known URLLC arrival scenario, the arrival of URLLC is deterministic, so they will not cause block error.
\section{Conclusion}
We developed a framework for joint eMBB and URLLC time-frequency resource allocation aimed to allocate radio resource to URLLC traffic with the characteristic of random arrival and optimize the throughput of eMBB users simultaneously. We proposed a method to predict the URLLC packets arrival and to configure eMBB coding redundancy. Our simulation results shows that the proposed methods can achieve $98\%$ throughput of the upper bound and higher probability of successful transmission in one-shot transmission than the conventional BWP method.

\bibliographystyle{IEEEtran}
\bibliography{ref}

\end{document}